\documentclass[journal]{IEEEtran}
\usepackage[latin1]{inputenc}
\usepackage{times,amsmath}
\usepackage{amssymb}
\usepackage{pstool}
\usepackage{subfigure}
\usepackage{multirow}
\usepackage{enumerate}
\usepackage{graphicx}
\usepackage{MnSymbol}
\usepackage{stfloats} 
\usepackage[table]{xcolor}
\usepackage[square, comma, sort&compress, numbers]{natbib}
\usepackage{nohyperref}
\usepackage{algorithm,algorithmic}
\usepackage{epstopdf}
\usepackage{lipsum}
\usepackage{mathtools}
\usepackage{cuted}
\graphicspath{ {Figures/} }
\usepackage[margin=1.5cm]{geometry}
\usepackage{multicol}

\begin{document}
\title{Performance Analysis of Underwater Acoustic Channel Amid Jamming by Random Jammers}

\author{
Waqas Aman, Saif Al-Kuwari, and Marwa Qaraqe

\thanks{{Authors gratefully acknowledge the G5828 grant by NATO's Science for Peace and Security Programme. \\ 
All the authors are with the Division of Information and Computing Technology, College of Science and Engineering, Hamad Bin Khalifa University, Qatar Foundation, Doha, Qatar.\\ Emails: \{waman, smalkuwari, mqaraqe\}@hbku.edu.qa}}
\vspace{-0.8cm}

}

\maketitle

\begin{abstract} 
Underwater communication networks are increasingly popularized by various important maritime applications. However, this also leads to an increased threat landscape. This letter presents the first study that considers jamming attacks by random jammers present in the surroundings of legitimate transceivers in underwater acoustic communication systems. We investigate the impact of jamming attacks on various performance parameters of the legitimate underwater acoustic communication link. In particular, we investigate the legitimate link using stochastic geometry for important performance parameters, namely coverage probability, average rate, and energy efficiency of the link between two legitimate nodes, i.e., underwater and surface nodes. We then derive and present tractable expressions for these performance parameters. Finally, we performed a Monte Carlo simulation to validate our analysis. We plot the performance metrics against the transmit power, and jamming power for different intensities of the jammers in shallow, mid, and deep water scenarios.  Results reveal that on average, jamming in deep water has a relatively high impact on the performance of legitimate link than in shallow water. 
\end{abstract}


\section{Introduction}
\label{sec:intro}
Recently, a considerable amount of research has been conducted to explore innovative approaches, from design to signal processing, on underwater acoustic communication \cite{fattah2020survey, cuiintegrated}. 
However, the broadcast nature of underwater acoustic communication makes it vulnerable to many types of malicious attacks \cite{aman2023security}. One of the prominent attacks is the jamming attack, which can lead to denial of services or high error rate in ongoing communication. In fact, as underwater acoustic communication systems operate in a limited frequency band due to the harsh environment, jamming can aggressively consume the available bandwidth, making it difficult for legitimate communication signals to pass through. On the other hand, resisting the jamming attack in the typical underwater environment is often difficult as it is increasingly challenging to physically access such hostile environments. This consequently makes it difficult to locate and remove jamming devices there. Therefore, investigating the potential impact of jamming in these environments is of utmost importance. 


The authors in \cite{misra2012jamming} consider a jamming attack scenario in the underwater acoustic communication network, and propose a detection scheme that detects jamming by observing abnormalities in the received data. Specifically, jamming is detected by measuring any abnormality in terms of packet transmission ratio or the amount of energy consumption. Once a jamming attack has been identified, the node sends
a high-priority packet to its neighbors and increases sleeping time to save energy. Meanwhile, it also maps a jammed
area to prevent data transmission from occurring, which mitigates the jamming effect. Similarly, the authors in \cite{xiao2018anti} propose a deep Q-network-based resource allocation mechanism that utilizes the transmit power of the legitimate transmitter node and the location of the receiver node to mitigate the jamming effect by enhancing the signal-to-jamming power ratio. More recently, the authors in \cite{10295387} utilize a double deep Q network (D2QN) to propose a mechanism that jointly optimizes the transmit power and receiver node' trajectory to maximize the achievable end-to-end throughput amid jamming.

Alternatively, game theory  
\footnote{{Game theory is a branch of mathematics that studies how entities (referred to as ``players") make strategic decisions in situations where the outcome of their choice depends not only on their actions but also on the actions of others.}} has been utilized to detect and mitigate jamming in underwater acoustic communication systems \cite{xiao2015jamming,chiariotti2020underwater,signori2020game, signori2021geometry, wang2022jamming}.
Game theory and reinforcement learning are jointly used
to detect jamming attacks by formulating the interactions between underwater nodes and attackers (i.e., players) as an underwater jamming game in \cite{xiao2015jamming}. The players choose their transmit power levels to maximize their individual utilities based on the SINR of the normal signals and
transmission costs. Nash equilibrium of a static jamming game is presented in a closed-form expression for the jamming scenario with known acoustic channel gains. For unknown dynamic underwater environments, a Q-learning-based antijamming method is proposed where each node chooses its transmit power with no information on the channel gain of the jamming attackers available. In addition, other work uses the game theory approach for various resource allocation problems during jamming \cite{chiariotti2020underwater,signori2020game, signori2021geometry,wang2022jamming}.

To the best of our knowledge, the impact of random jammers on the performance of the legitimate node has not been studied in underwater acoustic communication networks. Earlier work \cite{misra2012jamming, xiao2018anti, 10295387, xiao2015jamming,chiariotti2020underwater,signori2020game, signori2021geometry,wang2022jamming} considered mainly three node-setups (Alice, Bob, and a Jammer) or fixed geometry for jammers and proposed their techniques to counter jamming via resource allocation schemes through game theory or other optimization techniques. In this work, instead, we consider a more realistic scenario where we assume multi-random jammers in the surrounding area of the legitimate nodes, and whose continuous jamming in the operating frequency range affects the ongoing communication between legitimate nodes. We investigate (using stochastic geometry) how randomly distributed jammers in the seabed can impact the coverage, average rate (AR), and energy efficiency (EE) of the legitimate communication link.

Despite being the first work of its kind in the literature, this work is important from two different aspects. Firstly, it tackles a realistic scenario where the precise number and locations of jammers are unknown, introducing a nuanced perspective to the existing literature. Secondly, it underscores the crucial need for understanding system behavior in dynamic, stochastic environments, empowering design engineers to optimize system parameters effectively to achieve predefined performance metrics such as coverage probability, spectral efficiency, and energy efficiency in a jamming environment.

The rest of this letter is organized as follows: Section \ref{sec:sys-model} presents the system and signal model, and  Section \ref{sec:method} presents the performance analysis. We evaluate our proposals using simulation and present the results in Section \ref{sec:results}. Finally, the paper concludes in Section \ref{sec:conclusion} with a few concluding remarks and future directions.

\section{System \& Signal Model}
\label{sec:sys-model}
\begin{figure}
    \centering
    \includegraphics[width=0.5\textwidth]{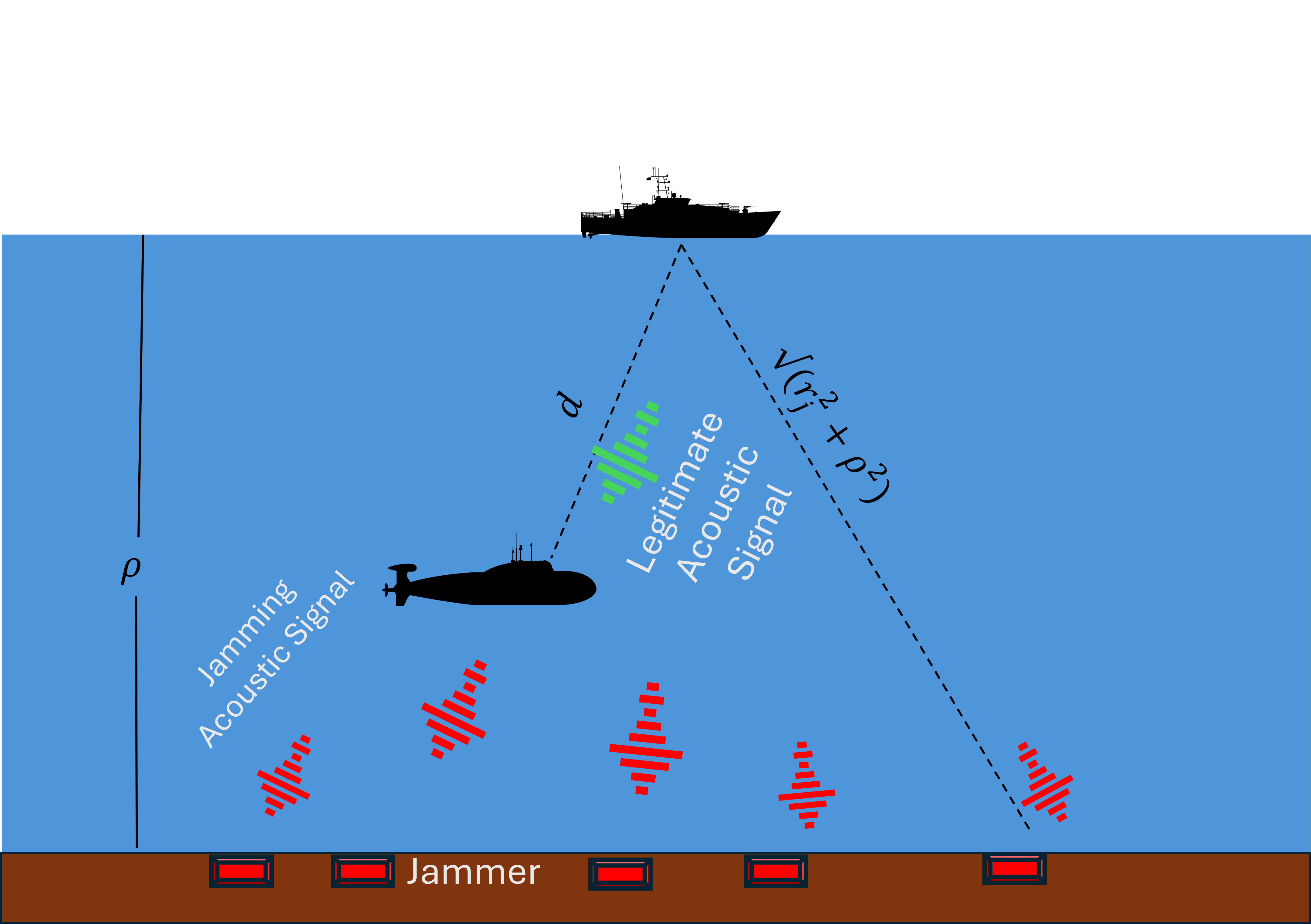}
    \caption{System Model}
    \label{fig:sm}
\end{figure}

We consider an underwater acoustic communication system as depicted in Fig. \ref{fig:sm} where we have a legitimate underwater transmitting node/submarine at a certain location that transmits signals at power $P_t$, and a surface receiver node/ship to receive and decode information. Furthermore,  
we consider random jammers at the seabed with a certain depth $\rho$ which are distributed according to a homogeneous Poisson point process (PPP) with intensity $\lambda_J$. All jammer nodes emit continuous jamming signals with power $P_J$ in the operating frequency band, effectively blocking or interfering with legitimate communication. 
Therefore, we are interested in finding how these random jammers, on average, affect the performance of legitimate communication. 

The clean received signal $y$ in the frequency domain of the legitimate transmitter on the surface ship is given as

\begin{align}
    y(f)=\sqrt{\frac{P_t}{\text{PL}(f)}}H(f)x+N(f),
\end{align}
where $P_t$ is the transmit power, $\text{PL}$ is pathloss, $x$ is the transmitted symbol, $H(f)$ is the channel transfer function, and $N(f)$ is PSD of the acoustic noise, can be expressed in dB as $N(f)_{\text{dB}} \approx N_1- \tau 10\log (f)$, where $N_1$ and  $\tau$ are the experimental constants. 
The path-loss $\text{PL}$ in dB of underwater acoustic (UWA) channel between a node pair having distance $d$ separation is given as \cite{qarabaqi2013statistical}


\begin{align}
\label{eq:pldb}
\text{PL}_{\text{dB}} (f,d)= \nu 10\log d + d \alpha(f)_{\text{dB}}. 
\end{align}

Eventually, the path-loss of UWA channel is the summation of spreading loss ($\nu 10\log d$) and absorption loss ($d \alpha(f)_{\text{dB}}$) in dB scale, where $\nu$ is the spreading factor, while  the  absorption coefficient $\alpha(f)_{\text{dB}}$ is given as $\alpha(f)_{\text{dB}}=\frac{0.11            
f^2}{1+ f^2}+\frac{44f^2}{4100+f^2}+2.75\times 10^{-4}f^2+0.003$ \cite{qarabaqi2013statistical}.




According to the statistical behavior of the channel transfer function \cite{stojanovic2007relationship}, it can be expressed as $H(f) \sim CN(\sum_{l=1}^Lc_lE\{h_l\},\sigma_L^2\sum_{l=1}^L\vert c_l\vert )$ where $c_l= e^{-j2\pi f\xi_l}$ with $\xi_l$ is the $l$-path delay, and $\sigma_L = \sigma_l \ \ \forall l$ and $L$ is the total number of paths. Next, $\vert H(f)\vert$ is distributed as Rician with shape parameter $K=\frac{\vert\sum_{l=1}^Lc_lE\{h_l\}\vert^2}{\sigma_L^2\sum_{l=1}^L\vert c_l\vert }=\frac{\vert\sum_{l=1}^Lc_lE\{h_l\}\vert^2}{\sigma_L^2 L}$. Equivalently, $\vert H(f)\vert \sim \text{Rice}(\sqrt{\frac{2}{L}}\mid \sum_{l=1}^Lc_lE\{h_l\}\mid , \sigma_L)$. Assuming unit variance of path's distribution (i.e $\sigma_l^2 = 1$), and omitting $f$ for the sake of brevity, $\vert H\vert^2$ is distributed as non-central chi-squared with two degrees-of-freedom and non-centrality parameter $\psi=\left( \sqrt{\frac{2}{L}} \mid \sum_{l=1}^Lc_lE\{h_l\}\mid \right)^2$. 
At this stage, assuming a narrow-band system \cite{stojanovic2007relationship},
we can write the signal-to-jamming and noise ratio (SJNR) as
\begin{align}
    \text{SJNR}=\frac{\zeta}{ J+\sigma^2},
\end{align}
where $\zeta =\frac{p_t}{\text{PL}}\vert H\vert^2$ is the received signal power from the legitimate transmitter, $J=  P_J \sum_{j\in \Phi_j} \frac{1}{\text{PL}_j}\vert H_j\vert^2$ is the aggregated jamming power with $P_J$ the jamming power, $\text{PL}_j$ is the path loss and $H_j$ is the channel transfer function of $j$-th jammer and $\sigma^2=\Delta f N(f)$ is the total noise power in the operating band $\Delta f$.

\section{Performance Analysis}
\label{sec:method}
We analyzed our system in terms of three metrics: coverage probability, AR, and EE. We derive Laplace transforms of received and jamming signals, which contribute to the three metrics.  We discuss each of them in detail below.

\subsection{Coverage Probability}
By definition, the conditional (i.e., conditioned on the distance of legitimate transmitter to surface node) coverage probability in the case of jamming can be expressed as
\begin{align}
  &  P_C \mid (D=d) =\text{P}[\text{SJNR}\mid (D=d) \geq \tau] = \text{P}[\frac{\frac{P_t\vert H \vert^2 }{\text{PL}(d)}} {J+\sigma^2}\geq \tau] \\
&    = \text{P}[{ \vert H \vert^2}\geq \frac{\text{PL}(d)}{P_t} \tau (J+\sigma^2)], \nonumber
\end{align}
where $D$ denotes an R.V. for the distance between legitimate underwater and surface nodes, $\tau$ is the threshold or target SJNR. 
To find this probability, we invoke the Gill-Paliez inversion theorem \cite{di2014stochastic}, accordingly the above equation can be written as
\begin{align}
\label{eq:PC_d}
    &P_C \mid (D=d)= \\
    &0.5-2\lambda_J\int_0^\infty \frac{\text{Img}[\mathcal{L}_{\vert H \vert^2}(-is\frac{P_t}{\text{PL}(d)\tau}){F}_{J}(is)]}{s} d_s, \nonumber
\end{align}
where Img [.] represents the imaginary part of a complex value function, $F_J(is)=\int_0^\infty y \exp{(iys\frac{\sigma^2}{P_J})} \exp{(-\pi \lambda_J y^2 \gamma(isy))}dy$ with $\gamma(isy)=\mathbf{E}[{}_1F_1(-2,-1,is \frac{P_J}{\text{PL}_j(y)}\vert H_j \vert^2)]$ where ${}_1F_1(.,.,.)$ is confluent hyper-geometric function of the first kind. The above probability is conditioned on distance $d$, now, the overall coverage probability $\text{P}_C$ can be computed by computing the expectation of Eq. \ref{eq:PC_d} w.r.t. to distance $D$, the final expression is presented in Eq. \ref{eq:PC}, where $d_{\text{max}}$ is the maximum value of R.V. $D$ (distance of the legitimate node). We consider R.V. $D$ as uniform R.V. distributed as $D \sim \mathcal{U}(0,d_{\text{max}})$. Thus, it implies, $f_D(d)=\frac{1}{d_{\text{max}}}$

\begin{strip}
\begin{equation}
\label{eq:PC}
    P_C=\int_0^{d_{\text{max}}}
    (0.5-2\lambda_J\int_0^\infty \frac{\text{Img}[\mathcal{L}_{\vert H \vert^2}(-is\frac{P_t}{\text{PL}(d)\tau}){F}_{J}(is)]}{s} d_s)f_{D}(d)d_{d},
\end{equation}

\begin{equation}
\label{eq:L_J}
  \mathcal{L}_{J}(s) = \exp{(-2\pi \lambda_J \int_0^\infty (1-\exp{(\frac{-\psi s P_j \frac{1}{\text{PL}_j(r)}}{1+2sP_j \frac{1}{\text{PL}_j(r)}})}(1+2sP_j \frac{1}{\text{PL}_j(r)})^{-1} )rd_r )}
\end{equation}

 \begin{equation}
 \label{eq:ASR_F}
   \text{AR}= \int_0^{d_{\text{max}}}\int_0^\infty \frac{\frac{(1-\exp{(\frac{-\hat{\psi}(d) s}{1+2s})})}{(1+2s)} \exp{(-2\pi \lambda_J \int_0^\infty (1-\exp{(\frac{-\hat{\psi}(d) s P_j \frac{1}{\text{PL}_j(r)}}{1+2sP_j \frac{1}{\text{PL}_j(r)}})}(1+2sP_j \frac{1}{\text{PL}_j(r)})^{-1} )rd_r )}}{s} \exp{(-\sigma^2s)} d_sf_D(d)d_d,
\end{equation} 
\end{strip}

\subsection{Average Rate}
 AR provides a more realistic assessment of a wireless communication system's performance in a time-varying and fading channel. Instead of looking at the instantaneous rate, which can fluctuate greatly, it considers the average rate over a long time period, which is more relevant for practical applications. By definition, it can be expressed as

\begin{align}
    \text{AR}=\mathbf{E} \left[ \log_2(1+\frac{\zeta}{ J+\sigma^2})  \right]  \ \ \ \text{bps},
\end{align}
where the expectation is joint w.r.t. $\zeta$ and $J$.
We invoke Hamdi's lemma to find the conditional AR, given as \cite{hamdi2010useful}


\begin{align}
\label{eq:ASR}
   \text{AR} \mid (D=d)= \int_0^\infty \frac{\mathcal{L}_{J}(s) -\mathcal{L}_{\zeta(d)}(s)\mathcal{L}_{J} (s)}{s}\exp{(-\sigma^2s)} d_s,
\end{align}
By putting Eqs. \ref{eq:L_J} \& \ref{eq:L_zeta} into Eq. \ref{eq:ASR}, we get the final expression for AR which is presented in Eq. \ref{eq:ASR_F}.

\subsection{Energy Efficiency}

Energy efficiency (EE) is an important consideration in underwater wireless communication. Due to the remote and harsh environment of underwater wireless communication, energy becomes a precious resource of the system due to expensive energy source replacement mechanisms. By reducing the energy consumption during communication, the operational costs associated with battery replacement and maintenance can be minimized. 
 When a jamming attack occurs, the device may need to increase its transmit power to overcome the interference \cite{Gan:ITL:2022}. Energy-efficient communication can help ensure that the device's battery lasts longer, allowing it to continue functioning even in the presence of jamming. The EE can be computed as:  
\begin{align}
   \text{EE} = \frac{\int_0^{d_{max}}\int_0^\infty \frac{\mathcal{L}_{J}(s) -\mathcal{L}_{\zeta(d)}(s)\mathcal{L}_{J} (s)}{s}\exp{(-\sigma^2s)} d_sd_d}{S_p+P_t},
\end{align}
where $S_p$ is the static power consumption.
\subsection{Laplace Transforms}
The Laplace transform (LT) of the received signal can be expressed as:
\begin{align}
\label{eq:L_zeta}
\mathcal{L}_{\zeta}(s) & = \mathbf{E}\left[ \exp{(-s\zeta)} \right] \\ 
    & \mathcal{L}_\zeta (s)=\exp{(\frac{-\hat{\psi} s}{1+2s})}(1+2s)^{-1}, \nonumber
\end{align}
where $\hat{\psi}=\left( \sqrt{\frac{2\text{PL}}{P_tL}} \mid \sum_{l=1}^Lc_lE\{h_l\}\mid \right)^2$ is the non-central parameter of non-central Chi-squared R.V. $\zeta$. Note that the LT of $\vert H \vert^2$ is the same as Eq. \ref{eq:L_zeta} but with a different non-central parameter $\psi$.
Similarly, the LT of jamming can be computed as:

\begin{align}
\label{LT_J}
        \mathcal{L}_{J}(s) & = \mathbf{E}\left[ \exp{(-sJ)} \right] \\ 
      & \stackrel{(a)} =\mathbf{E}_{\Phi,\{\vert H_j \vert^2 \}} \left[ \exp{(-s P_j \sum_{j\in \Phi_j} \frac{1}{\text{PL}_j(\Phi)}\vert H_j \vert^2 )} \right] \nonumber \\
      & \stackrel{(b)}= \mathbf{E}_{\Phi},\mathbf{E}_{\{\vert H_j \vert^2 \}} \left[ \prod_j \exp{(-s P_j \frac{1}{\text{PL}_j(\Phi)}\vert H_j \vert^2 )} \right] \nonumber \\
       & \stackrel{(c)}= \mathbf{E}_{\Phi} \left[ \prod_j \exp{(\frac{-\psi s P_j \frac{1}{\text{PL}_j(\Phi)}}{1+2sP_j \frac{1}{\text{PL}_j(\Phi)}})}(1+2sP_j \frac{1}{\text{PL}_j(\Phi)})^{-1} \right] \nonumber 
\end{align}


 Eq. \ref{LT_J} (a) expresses the main random variable into its full form. Eq. \ref{LT_J} (b) is obtained by using the exponential function property and the fact that the fading gains $\vert H_j \vert^2$ and the point process are independent, and Eq. \ref{LT_J} (c) is due to i.i.d of $H_j \ \forall j$. Finally, using probability generating functional (PGFL) for the PPP, the final expression for the LT of jamming is presented in Eq. \ref{eq:L_J},
 where $\text{PL}_j(r)(f)[dB]= \nu 10\log (\sqrt{r^2+\rho^2}) + \sqrt{r^2+\rho^2}\alpha(f)_{\text{dB}}.$ is the path loss in dB of j-th jammer to the surface node, $r$ is the random 2D distance and $\rho$ is the distance/depth from the surface to the seabed.
\color{black}
\section{Simulation results}
\label{sec:results}
We use Matlab to develop the simulations presented in this section. The simulation parameters are taken from commercially available Popoto acoustic modems \cite{popoto}. Unless specified otherwise, we set static power consumption to $S_p=1.5$ Watts, transmit power $P_t=20$ Watts, jamming power $P_J=20$ Watts, operating frequency $f=22$ KHz, bandwidth $10$ KHz, $\rho=0.1, 1 $ and $2$km for shallow, mid and deep water respectively \cite{Jason:water}, $d_{max}=\sqrt{10^2+\rho^2}$ km, and $\tau=2$. Furthermore, to validate the analysis, Monte Carlo results are also plotted where $10^6$ total number of node's deployment is taken for computing the performance metrics. We use an acoustic channel simulator for underwater communication to generate channel gains for Monte Carlo results \cite{qarabaqi2013statistical}. 

\begin{figure}
    \centering
    \includegraphics[scale=0.5]{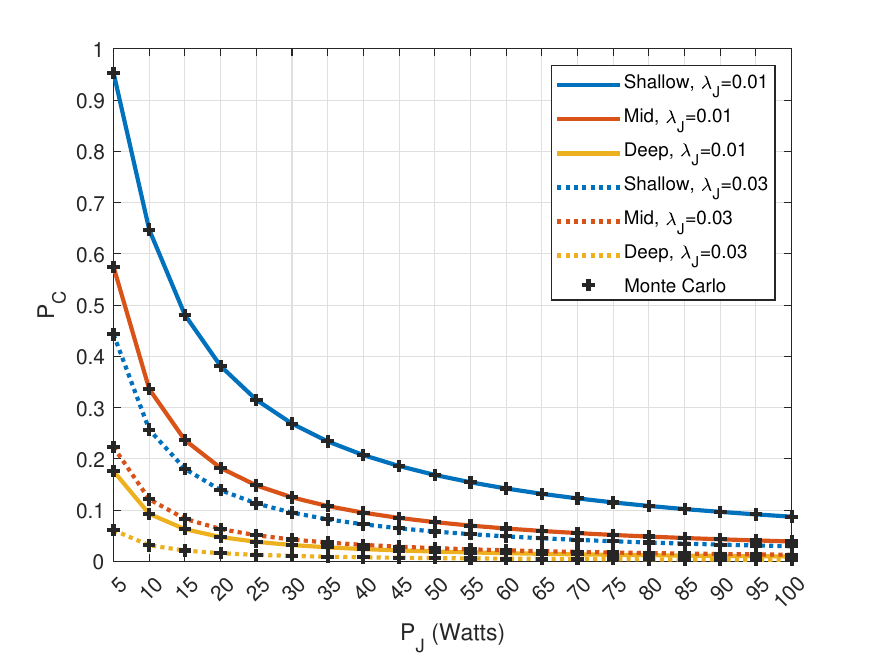}
    \caption{Coverage probability against jamming power of jammers}
    \label{fig:CP_v_PJ}
\end{figure}

\begin{figure}
    \centering
    \includegraphics[scale=0.5]{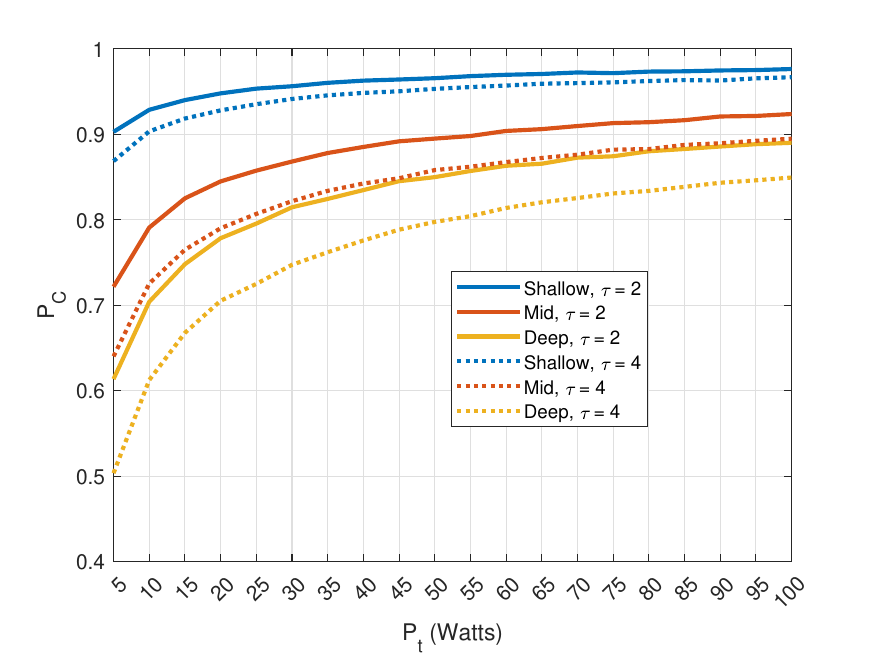}
    \caption{Coverage probability against transmit power of legitimate node}
    \label{fig:CP_v_tau}
\end{figure}

Fig. \ref{fig:CP_v_PJ} shows the performance of overall coverage probability against jamming power. In this figure, we chose two values for the average number of jammers, $\lambda_j=0.01$ (ten jammers per km) and $\lambda_j=0.03$ (thirty jammers per km). We observe that overall coverage probability decreases significantly when an increase occurs in the average number of jammers. This is obvious, more jammers mean more jamming power and hence enhanced interference from the jammers into legitimate communication, thus reducing overall coverage probability. We also observe that overall coverage probability is relatively higher in shallow water than in mid and deep water. This is because on average the legitimate node in mid and deep-water cases is farther away from the surface node and thus has a relatively more degraded channel. We also plot the Monte Carlo results and observe a perfect match with analytical results. Fig. \ref{fig:CP_v_tau} shows the overall coverage probability against the transmit power of a legitimate node for two values (2 \& 4) of target SJNR $\tau$. We observe an increase in the probability with the increase in the transmit power. Here, one can tell about the percentage of time one can get a target SJNR value, for example, we observe that for shallow water almost 98$\%$ of the time one will meet the $\tau=2$ for $\lambda_J =0.03$. Further, a similar trend for shallow, mid, and deep water to Fig. \ref{fig:CP_v_PJ} is observed.

\begin{figure}[htb!]
    \centering
    \includegraphics[scale=0.5]{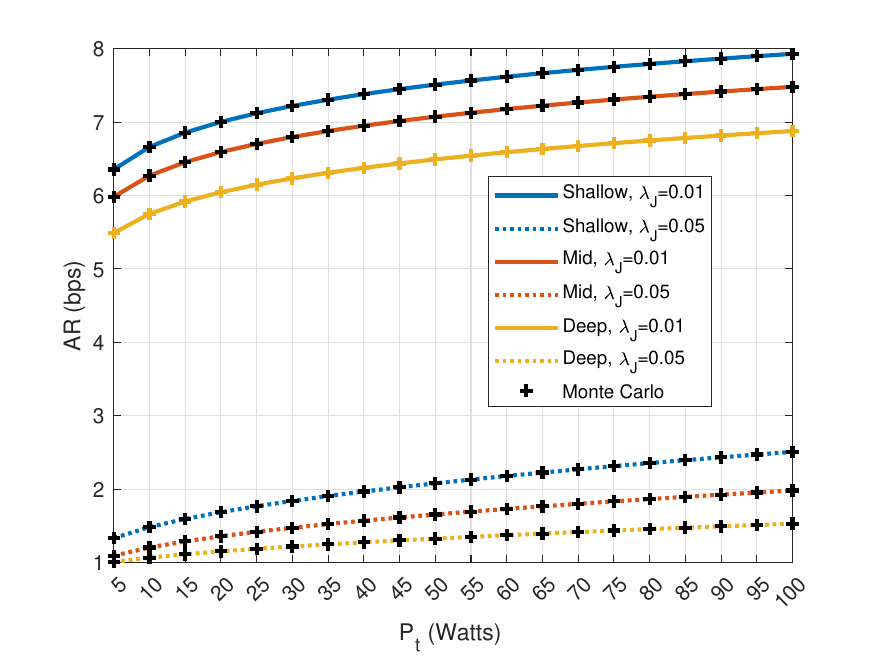}
    \caption{Average rate (AR) in \textit{bps} against transmit power of legitimate node}
    \label{fig:AR_v_Pt}
\end{figure}

\begin{figure}[htb!]
    \centering
    \includegraphics[scale=0.5]{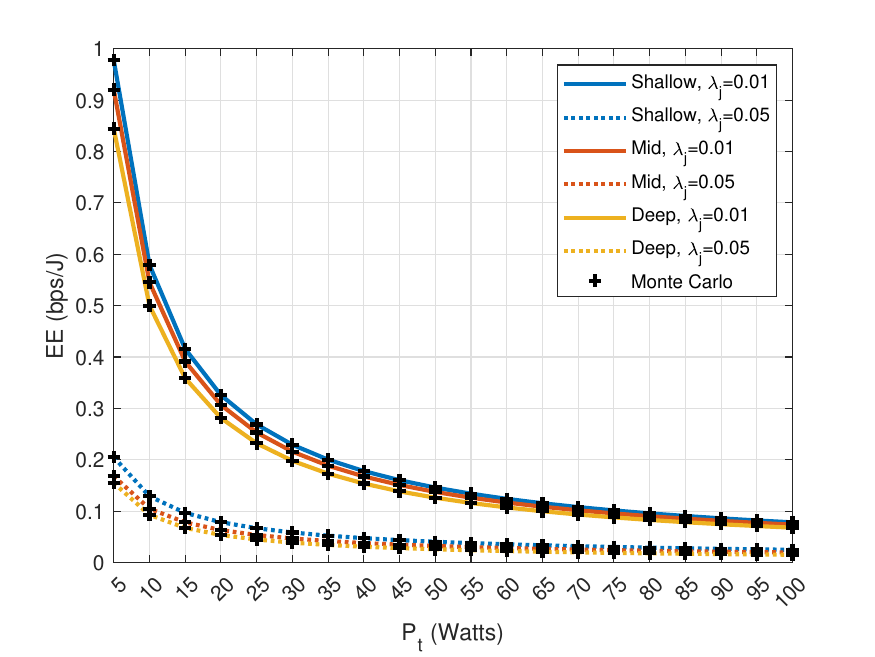}
    \caption{Energy efficiency (EE) in \textit{bps/Joule} against transmit power of legitimate node}
    \label{fig:EE_v_Pt}
\end{figure}
\begin{figure}[htb!]
    \centering
    \includegraphics[scale=0.5]{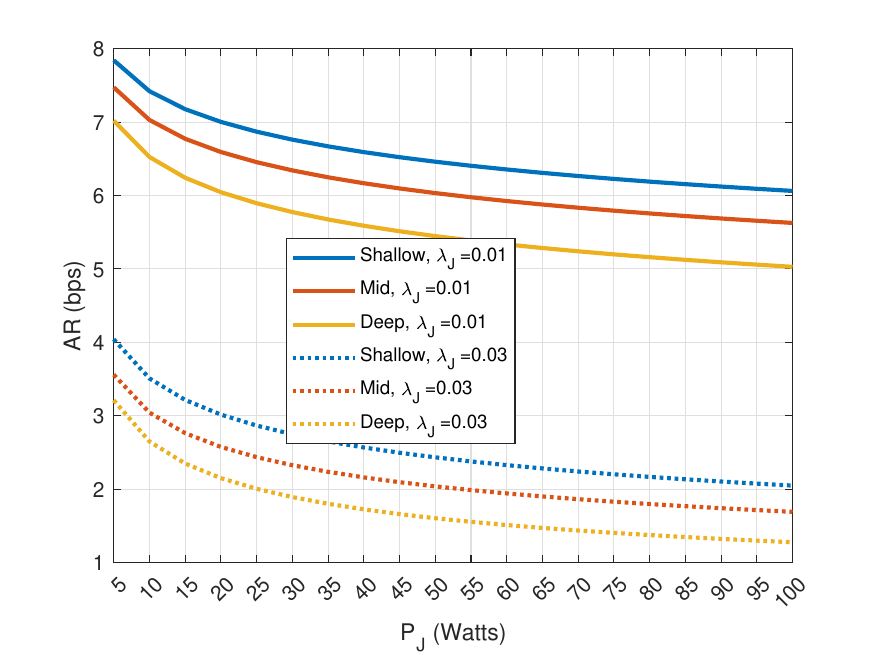}
    \caption{Average rate (AR) in \textit{bps} against jamming power}
    \label{fig:AR_v_PJ}
\end{figure}
\begin{figure}[htb!]
    \centering
    \includegraphics[scale=0.5]{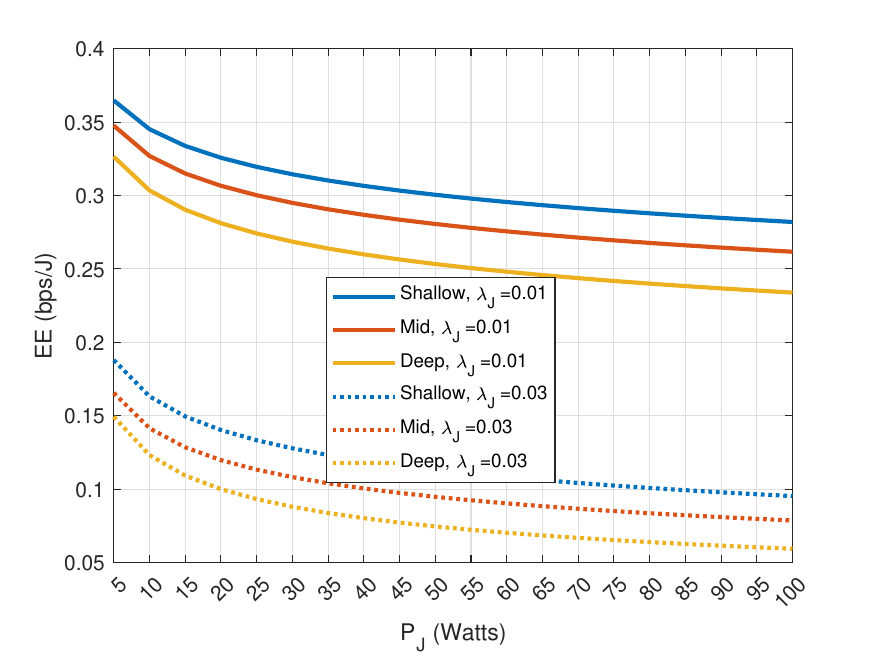}
    \caption{Energy efficiency (EE) in \textit{bps/Joule} against jamming power}
    \label{fig:EE_v_PJ}
\end{figure}

In Fig. \ref{fig:AR_v_Pt}, we keep the jamming power of all the jammers fixed to the maximum value (i.e., $20$ Watts) according to the commercially available modems, and assume a scenario where the legitimate transmitter is cable of varying its transmit power over a range $5$ to $100$ Watts. We observe that increasing transmit power increases the average rate, and shallow water has a higher average rate than mid and deep-water scenarios. In Fig. \ref{fig:EE_v_Pt}, we plot energy efficiency against transmit power for the same configuration as Fig. \ref{fig:AR_v_Pt}. We observe that on the one hand in Fig. \ref{fig:AR_v_Pt}, the transmit power enhances the average rate but on the other side it reduces the energy efficiency of the system as shown in Fig. \ref{fig:EE_v_Pt}. So, this implies there is a tradeoff between the two, one needs to comprise one for the other. The important thing to note here is that increasing transmit power increases the average rate but with a low slope while the descending in the EE plot is relatively high, thus making transmit power a vital system parameter to be considered. We also plot the Monte Carlo results, their perfect match with analytical results validate the analysis. 

Next, we assume a scenario where the legitimate node is transmitting at a fixed power and the jammers are capable of varying their jamming powers over a range $5$ to $100$ Watts. We can see that in both Figs. \ref{fig:AR_v_PJ} \& \ref{fig:EE_v_PJ}, the trend of AR and EE is decreasing against the increase in jamming power. This is due to the fact that more jamming power means more interference and thus low average rate and EE. This means that increasing jamming power does not go in favor of either of both.  
\section{Conclusion}
\label{sec:conclusion}
In this letter, we investigated jamming attacks by random jammers in an underwater acoustic communication system. We utilized stochastic geometry to evaluate the performance of legitimate communications in terms of overall coverage probability, average rate, and energy efficiency. We provided tractable expressions for all three aforementioned performance metrics. To validate the analysis the analytical results are compared against Monte Carlo results. 

This work is an initial attempt to averagely evaluate a legitimate underwater acoustic link that is under the influence of jamming by randomly located jammers. Looking forward, a multitude of avenues beckon for further exploration. One promising trajectory involves delving into optimal resource allocation strategies aimed at enhancing coverage, average rate, and energy efficiency. Additionally, considering the ever-present challenge posed by random blockages such as aquatic life or submarines, future investigations could delve into the implications of these obstructions on system performance, thereby enriching our understanding of real-world operational dynamics.





\footnotesize{
\bibliographystyle{IEEEtran}
\bibliography{references}

\begin{thebibliography}{10}
\providecommand{\url}[1]{#1}
\csname url@rmstyle\endcsname
\providecommand{\newblock}{\relax}
\providecommand{\bibinfo}[2]{#2}
\providecommand\BIBentrySTDinterwordspacing{\spaceskip=0pt\relax}
\providecommand\BIBentryALTinterwordstretchfactor{4}
\providecommand\BIBentryALTinterwordspacing{\spaceskip=\fontdimen2\font plus
\BIBentryALTinterwordstretchfactor\fontdimen3\font minus \fontdimen4\font\relax}
\providecommand\BIBforeignlanguage[2]{{%
\expandafter\ifx\csname l@#1\endcsname\relax
\typeout{** WARNING: IEEEtran.bst: No hyphenation pattern has been}%
\typeout{** loaded for the language `#1'. Using the pattern for}%
\typeout{** the default language instead.}%
\else
\language=\csname l@#1\endcsname
\fi
#2}}

\bibitem{fattah2020survey}
S.~Fattah, A.~Gani, I.~Ahmedy, M.~Y.~I. Idris, and I.~A. Targio~Hashem, ``A survey on underwater wireless sensor networks: Requirements, taxonomy, recent advances, and open research challenges,'' \emph{Sensors}, vol.~20, no.~18, p. 5393, 2020.

\bibitem{cuiintegrated}
X.~Cui, Y.~Cai, J.~Li, L.~Li, B.~Jiang, and L.~Liu, ``An integrated waveform design method for underwater acoustic detection and communication,'' \emph{Internet Technology Letters}, p. e522.

\bibitem{aman2023security}
W.~Aman, S.~Al-Kuwari, M.~Muzzammil, M.~M.~U. Rahman, and A.~Kumar, ``Security of underwater and air--water wireless communication: State-of-the-art, challenges and outlook,'' \emph{Ad Hoc Networks}, vol. 142, p. 103114, 2023.

\bibitem{misra2012jamming}
S.~Misra, S.~Dash, M.~Khatua, A.~V. Vasilakos, and M.~S. Obaidat, ``Jamming in underwater sensor networks: detection and mitigation,'' \emph{IET communications}, vol.~6, no.~14, pp. 2178--2188, 2012.

\bibitem{xiao2018anti}
L.~Xiao, X.~Wan, W.~Su, Y.~Tang, \emph{et~al.}, ``Anti-jamming underwater transmission with mobility and learning,'' \emph{IEEE Communications Letters}, vol.~22, no.~3, pp. 542--545, 2018.

\bibitem{10295387}
H.~Zhang, L.~Wu, Y.~Zhi, C.~Yang, X.~Cao, J.~Zhang, and H.~Li, ``Throughput maximization for usv-enabled underwater wireless networks under jamming attack,'' \emph{IEEE Sensors Journal}, pp. 1--1, 2023.

\bibitem{xiao2015jamming}
L.~Xiao, Q.~Li, T.~Chen, E.~Cheng, and H.~Dai, ``Jamming games in underwater sensor networks with reinforcement learning,'' in \emph{2015 IEEE Global Communications Conference (GLOBECOM)}.\hskip 1em plus 0.5em minus 0.4em\relax IEEE, 2015, pp. 1--6.

\bibitem{chiariotti2020underwater}
F.~Chiariotti, A.~Signori, F.~Campagnaro, and M.~Zorzi, ``Underwater jamming attacks as incomplete information games,'' in \emph{IEEE INFOCOM 2020-IEEE Conference on Computer Communications Workshops (INFOCOM WKSHPS)}.\hskip 1em plus 0.5em minus 0.4em\relax IEEE, 2020, pp. 1033--1038.

\bibitem{signori2020game}
A.~Signori, F.~Chiariotti, F.~Campagnaro, and M.~Zorzi, ``A game-theoretic and experimental analysis of energy-depleting underwater jamming attacks,'' \emph{IEEE Internet of Things Journal}, vol.~7, no.~10, pp. 9793--9804, 2020.

\bibitem{signori2021geometry}
A.~Signori, F.~Chiariotti, F.~Campagnaro, R.~Petroccia, K.~Pelekanakis, P.~Paglierani, J.~Alves, and M.~Zorzi, ``A geometry-based game theoretical model of blind and reactive underwater jamming,'' \emph{IEEE Transactions on Wireless Communications}, vol.~21, no.~6, pp. 3737--3751, 2021.

\bibitem{wang2022jamming}
H.~Wang, Y.~Huang, and X.~Zeng, ``Jamming games in underwater sensor networks with hierarchical learning,'' in \emph{2022 8th International Conference on Big Data Computing and Communications (BigCom)}.\hskip 1em plus 0.5em minus 0.4em\relax IEEE, 2022, pp. 428--434.

\bibitem{qarabaqi2013statistical}
P.~Qarabaqi and M.~Stojanovic, ``Statistical characterization and computationally efficient modeling of a class of underwater acoustic communication channels,'' \emph{IEEE Journal of Oceanic Engineering}, vol.~38, no.~4, pp. 701--717, 2013.

\bibitem{stojanovic2007relationship}
M.~Stojanovic, ``On the relationship between capacity and distance in an underwater acoustic communication channel,'' \emph{ACM SIGMOBILE Mobile Computing and Communications Review}, vol.~11, no.~4, pp. 34--43, 2007.

\bibitem{di2014stochastic}
M.~Di~Renzo and P.~Guan, ``Stochastic geometry modeling of coverage and rate of cellular networks using the gil-pelaez inversion theorem,'' \emph{IEEE Communications Letters}, vol.~18, no.~9, pp. 1575--1578, 2014.

\bibitem{hamdi2010useful}
K.~A. Hamdi, ``A useful lemma for capacity analysis of fading interference channels,'' \emph{IEEE Transactions on Communications}, vol.~58, no.~2, pp. 411--416, 2010.

\bibitem{Gan:ITL:2022}
\BIBentryALTinterwordspacing
R.~Gan, Y.~Wang, Z.~Xiao, H.~Zhang, X.~Wang, and M.~Yang, ``Energy-efficient dos attack against remote state estimation,'' \emph{Internet Technology Letters}, vol.~5, no.~4, p. e370, e370 ITL-21-0190.R1. [Online]. Available: \url{https://onlinelibrary.wiley.com/doi/abs/10.1002/itl2.370}
\BIBentrySTDinterwordspacing

\bibitem{popoto}
``{Popoto Modem},'' \url{https://www.popotomodem.com/}, 2020.

\bibitem{Jason:water}
J.~Lavis, ``{Shallow, mid to ultra deepwater definitions},'' \url{https://drillers.com/shallow-mid-to-ultra-deepwater-definitions/}.

\end{thebibliography}
}

\vfill\break

\end{document}